\documentclass[a4paper,12pt]{article}
\usepackage{amsmath}
\usepackage{amssymb}
\usepackage{latexsym}
\usepackage{mathrsfs}
\usepackage[dvips]{graphicx}
\usepackage{pstricks}
\usepackage{color}
\usepackage{amsthm}
\usepackage{thmtools}
\usepackage[colorlinks=true, citecolor=blue]{hyperref}

\def\f(#1){{\mathop{f}^{(#1)}}}
\def\m(#1){{\mathop{m}^{(#1)}}}
\def\C(#1){{\mathop{C}^{(#1)}}}
\def\p(#1){{\mathop{p}^{(#1)}}}

\def\ben{\begin{equation}}
\def\een{\end{equation}}
\def\bena{\begin{eqnarray}}
\def\eena{\end{eqnarray}}

\def\d{{\rm d}}

\def\C{{\cal C}}

\declaretheoremstyle[
spaceabove=6pt, spacebelow=6pt, headfont=\sc,
notefont=\mdseries, notebraces={(}{)},
bodyfont=\normalfont, postheadspace=1em,
qed=\qedsymbol
]{myproof}
\declaretheorem[name=Proof,unnumbered,style=myproof]{prf}
\declaretheorem[name=Theorem,thmbox=M]
{thm}

\begin{document}

\title{Non-existence of toroidal cohomogeneity-1 \\ near horizon geometries}

\author{
Jan Holland\thanks{\tt HollandJW1@Cardiff.ac.uk}\:
\\ \\
{\it School of Mathematics, Cardiff University} \\
{\it Cardiff, United Kingdom} \medskip \\
}

\date{August 3, 2010}

\maketitle

\begin{abstract}
We prove that $D\geq 5$ dimensional stationary, non-static near horizon geometries with $(D-3)$ rotational symmetries subject to the vacuum Einstein equations including a cosmological constant cannot have toroidal horizon topology. In $D=4$ dimensions the same result is obtained under the assumption of a non-negative cosmological constant.
\end{abstract}

\section{Introduction}

Extremal black holes, i.e. those with a degenerate Killing horizon (or equivalently vanishing surface gravity), have received increasing attention in recent years due to their special mathematical properties. Despite the fact that they are not believed to be physically realised in nature, extremal black holes have proven to be an interesting subject of study in the context of string theory, where it was possible to derive the Bekenstein-Hawking entropy formula \cite{David:2002wn}, as well as in supergravity theories in the presence of supersymmetries, where black holes are automatically extremal \cite{Gauntlett:2002nw}.

A particularly useful concept in the study of horizon properties of extremal black holes is their \emph{near horizon geometry}. As suggested by the name, the idea is to restrict attention to the immediate vicinity of the horizon and neglect any information on the remaining spacetime by performing a scaling process. This procedure strongly increases the tractability of many problems, but still recovers valuable physical information, e.g. in the context of the \emph{Kerr-CFT correspondence} \cite{Guica:2008mu,Hartman:2008pb,Compere:2009dp} . Rigorous definitions of the near horizon limit can be found in \cite{Reall:2002bh,Kunduri:2007vf}.

Recently, \emph{cohomogeneity-1} near horizon metrics have been classified in \cite{Kunduri:2008rs} (4 and 5 dimensions) and \cite{Hollands2009} (arbitrary dimension). Here cohomogeneity-1 means that additional rotational symmetries are assumed, such that the metric components only depend on a single parameter nontrivially\footnote{In a $D$-dimensional spacetime this requires $(D-3)$ additional commuting Killing vector fields}. In these derivations certain results relied crucially on the horizon topology, which is restricted to the cases \cite{kim1974torus,Hollands2008a}

\ben
H\cong \begin{cases} S^{3}\times T^{D-5} \\
S^{2}\times T^{D-4} \\
L(p,q)\times T^{D-5}\\
 T^{D-2} \quad ,
\end{cases}
\een
where $D$ is the spacetime dimension, $H$ a horizon cross section, i.e. a compact $(D-2)$-dimensional manifold with $U(1)^{D-3}$ action, $S^{n}$ the $n$-sphere, $T^{n}$ the $n$-torus and $L(p,q)$ Lens spaces.
 It turns out that the last case, i.e. \emph{toroidal} horizon topology, is somewhat special, both from a physical and purely mathematical viewpoint. This type of solution could not be classified by the methods of \cite{Kunduri:2008rs, Hollands2009} and was therefore excluded by hand,
  motivated by the fact that such a near horizon geometry could not arise as the scaling limit of a physical black hole spacetime due to the topological censorship theorem \cite{Friedman:1993ty, Chrusciel2008,Galloway2001}. In this paper we prove that a near horizon metric with toroidal horizon topology admitting the mentioned symmetries and satisfying the vacuum Einstein equations with 
  cosmological constant not only cannot arise as the mentioned scaling limit, but
  in fact does not exist. This closes a gap in the existing literature showing that the classification of \cite{Hollands2009} actually covers all possible solutions of the type considered.
  
 \paragraph{Notation:} In the following we use letters from the beginning of the roman alphabet, $a,b=1,\ldots,(D-2)$, as indices attached to quantities on the horizon cross section $H$ and letters from the middle of the roman alphabet, $i,j=1,\ldots, (D-3)$, as indices for quantities on an orbit generated by $U(1)^{D-3}$ on $H$.
  
 \section{Theorem and proof}

 Our general setting is as follows (see \cite{Hollands2009} for more details): We consider metrics of \emph{Gaussian null form} \cite{moncrief1983symmetries}

\ben
g=2\d v(\d u + u^{2}\alpha\d v+u \beta_{a}\d y^{a})+\gamma_{ab}\d y^{a}\d y^{b}
\label{metric0}
\een
on a $D$-dimensional manifold $M$, where $K=\partial/\partial v$ and $X=u\partial/\partial u-v\partial/\partial v$ are Killing vector fields and the function $\alpha$, one form $\beta=\beta_{a}\d y^{a}$ and two form $\gamma=\gamma_{ab}\d y^{a}\d y^{b}$ depend on neither $u$ nor $v$. $\gamma$ is actually a smooth metric on a compact manifold $H$ located at $u=v=0$ (the horizon cross section), and $\alpha$ and $\beta$ can be viewed as fields on $H$. This is the general form of the near horizon geometry obtained by introducing Gaussian null coordinates and taking the ''near horizon limit'', but it is not assumed that our metric arises from this procedure. In addition to $K$ and $X$ we assume $(D-3)$ commuting Killing fields $\psi_{1},\ldots, \psi_{D-3}$ tangent to $H$, generating the symmetry group $U(1)^{D-3}$ and also commuting with $K$ and $X$. Due to this increased isometry group our metric functions can only depend nontrivially on one single variable, so that this type of metric may be called ''cohomogeneity-one''. For non-toroidal topology of $H$ these solutions to the vacuum Einstein equations (with vanishing cosmological constant) have been classified by \cite{Hollands2009}. Following their analysis we will soon observe differences, and eventually a contradiction, in the toroidal case. Thus, the main result of this paper is the following:
\vspace{6pt}
\begin{thm}
There cannot be any smooth, stationary, non-static cohomogeneity-one near horizon geometry with topology $H\cong T^{D-2}$ satisfying the vacuum Einstein equations with
cosmological constant in $D\geq 5$ spacetime dimensions.
\end{thm}

\begin{prf}

We proceed in two steps: First we adapt the construction of suitable near horizon coordinates given in \cite{Hollands2009} to the case of toroidal horizon topology, which essentially only affects the coordinate $x$ and the 1-form $\beta$ (to be introduced below). The main difference to the non-toroidal case lies in the fact that here the orbit space $\hat{H}=H/U(1)^{D-3}$ is a circle, instead of a closed interval, and that the Gram matrix

\ben
f_{ij}=\gamma(\psi_{i},\psi_{j})
\een
is non-singular on all of $H$ \cite{Hollands2008a}.
  In the second step we will impose the vacuum Einstein equations on the resulting metric and show, assuming non-staticity and smoothness, that these cannot be fulfilled, proving our assertion.

\subsubsection*{Construction of coordinates}

We are going to define coordinates on $H$ adapted to our problem and express $\gamma$ in terms of these.
To begin with,  consider the 1-form
\ben
\Sigma=
*_{\gamma}(\tilde{\psi_{1}}\wedge\cdots\wedge\tilde\psi_{D-3})
\label{sigma}
\een 
on $H$, where $*_{\gamma}$ denotes the Hodge dual with respect to the metric $\gamma$ and where $\tilde\psi_{i}$
are the 1-forms obtained from the vector fields $\psi_i$ by lowering the index with $\gamma$.
The fact that the $\psi_{i}$ are commuting Killing fields implies that 
$\Sigma$
is closed and Lie-derived by all $\psi_{i}$. It can therefore be viewed as a closed 1-form on the orbit space $\hat{H}=H/U(1)^{(D-3)}$, which as mentioned above is just a circle, $T^{D-2}/U(1)^{(D-3)}= S^{1}$, in the case at hand. Now consider the vector field $\xi^{a}=\gamma^{ab}\Sigma_{b}/\det f$, which is well-defined because $\det f$ cannot vanish. It is easy to see that $\xi$ is orthogonal to the rotational Killing vector fields, i.e. $\gamma(\xi,\psi_{i})=0$. Thus, $\xi$ is normal to the orbits generated by the $\psi_{i}$, and we also have the relations $[\xi,\psi_{i}]=0$ and $[\psi_{i},\psi_{j}]=0$. By Frobenius' Theorem we can therefore introduce local coordinates $(x,\varphi^{1},\ldots, \varphi^{D-3})$ on $H$ with $\varphi^{i}\in[0,2\pi]$ such that
 \bena
 (\xi)^{a}&=&\left(\frac{\partial}{\partial x}\right)^{a}\\
 (\psi_{i})^{a}&=&\left(\frac{\partial}{\partial\varphi^{i}}\right)^{a}\quad .
 \eena
 In these coordinates our metric takes the form
 \ben
 \gamma=\frac{1}{\det f}\d x^{2} + f_{ij}\d\varphi^{i}\d\varphi^{j}
 \een
 and $\d x = \Sigma$, showing that $\d x$ (but not $x$ itself) is globally defined. 
The geometric significance of the coordinate $x$ is that it locally labels the orbits, i.e. it can 
 be alternatively viewed as a local coordinate on $\hat H \cong S^1$. It is defined up to the 
 period $P = \int_{\hat H} \Sigma$, which cannot be equal to zero due to the fact that $\Sigma$ is nowhere vanishing. The coordinates $\varphi^i$ are local coordinates on 
 each given orbit.\footnote{Thus, on $H$ we have the identifications
 \ben
 (x,\varphi^{1},\ldots,\varphi^{D-3})=(x+P,\varphi^{1}+\alpha^{1},\ldots,\varphi^{D-3}+\alpha^{D-3})
 \een
 where $\alpha^{i}$ are the angles that are obtained by starting from a given point labeled by $\varphi^i = 0$ on some orbit and following 
 the integral curve of $\xi$ until we come back to this orbit.}
 The periodicity of $x$ constitutes the first difference to the considerations of \cite{Hollands2009}.
 
Next consider the 1-form $\beta$, which we can decompose as 
$\beta=\beta_{x}(x)\d x+\beta_{i}(x)\d\varphi^{i}$ on $H$. As for any generic 1-form, we can further decompose  $\beta_{x} \d x$ into an exact- and a non-exact contribution:

\ben
\beta_{x}(x)\d x=\d\lambda+\frac{A}{
\det f}\d x\quad .
\een
Here $\lambda$ is a smooth function on $H$ that is Lie-derived by the $\psi_i$, or 
alternatively a function on $\hat H \cong S^1$. Integrating the equation over $H$ gives
\ben
\int_{S^{1}}\left(\beta_{x}(x)-\frac{A}{
\det f}\right)\d x=0
\een
which we may take as the definition of $A$.
Here we encounter the second major difference to the derivation in \cite{Hollands2009}: Whereas in the non-toroidal case it was possible to argue $A=0$ using the vanishing of the Gram determinant $\det f$ at the end points of the interval $H/U(1)^{D-3}$, 
we do not know this yet in the present case. Thus, our expression for $\beta$ will look slightly different (compare eq. (2.8) of \cite{Hollands2009}).

%

\begin{equation}
\beta=\d\lambda+\frac{A}{
\det f}\text{d}x+
e^{\lambda}k_{i}\text{d}\varphi^{i}
\end{equation}
where we have defined

\ben
k_{i}=
e^{-\lambda}\psi_{i}\cdot\beta\quad .
\een
The remaining two coordinates can be chosen as in \cite{Hollands2009}, so we keep $v$, define $r:= ue^{\lambda}$ and our metric takes the form

\ben
g=e^{-\lambda}[2\d v\d r+ B r^{2}\d v^{2}]+\frac{\d x^{2}}{
\det f}+f_{ij}(\d\varphi^{i}+
rk^{i}\d v)(\d\varphi^{j}+
rk^{j}\d v)+\frac{2re^{-\lambda}A}{
\det f} \d v\d x
\label{metric1}
\een
where $B=(2\alpha e^{-\lambda}-e^{\lambda}k_{i}k^{i})$. Note that this metric differs from the one given in eq.(2.11) of \cite{Hollands2009} only by the last term.

\subsubsection*{Employing Einstein's equations}

Having set up our coordinate system, we would like to impose the vacuum Einstein equations including a cosmological constant $\Lambda\in\mathbb{R}$. For a general metric of the form \eqref{metric0} these can be expressed as the following set of equations on $H$ (see e.g. \cite{Kunduri:2008rs})
\bena
R_{ab}&=&\frac{1}{2}\beta_{a}\beta_{b}-\nabla_{(a}\beta_{b)}+\Lambda\gamma_{ab} \label{einst1} \\
2\alpha&=&\frac{1}{2}\beta_{a}\beta^{a}-\frac{1}{2}\nabla_{a}\beta^{a}+\Lambda \label{einst2}
\eena
where $R_{ab}$ and $\nabla_{a}$ are the Ricci tensor and Levi-Civita connection associated to the horizon metric $\gamma$. Our strategy now follows three basic steps:
\begin{enumerate}
\item Show $k^{i}=const.$ using the $xi$-component of the Ricci tensor.
\item Show $B=const.$ using the contracted Bianchi identity for $R_{ab}$.
\item Derive a contradiction to our assumption of non-staticity from the $ij$-component of the Ricci tensor.
\end{enumerate}
To begin with, 
we obtain from the $xi$-component of eq.\eqref{einst1}

\begin{equation}
\partial_{x}k^{i}=\frac{A}{\det f}k^{i}\qquad\Rightarrow
k^{i}=K^{i}\exp{\left(\int_{0}^{x}\frac{A}{\det
f}\text{d}x'\right)}
\end{equation}
with some constants $K^{i}\in \mathbb{R}$. However, due to the toroidal horizon
topology, $x$ is a periodic coordinate as argued above, so
$k^{i}(x)=k^{i}(x+P)$. This implies $A=0$ or $K^{i}=0$ for all $i$, since otherwise
$k^{i}$ would be strictly monotonous in $x$ (the integrand $\frac{A}{\det f}$
always has the same sign). The latter just corresponds to a static solution \cite{Kunduri:2008rs, Hollands2009}, so it will not be considered here. Therefore,

\begin{equation}
A=0\quad \Rightarrow \quad k^{i}=\text{const}
\end{equation}
and $k^{i}\neq 0 $ for some $i$. Hence, as the last term in eq.\eqref{metric1} vanishes, our metric takes precisely the same form as in the non-toroidal case. Following \cite{Kunduri:2008rs}, it is next possible to deduce that $B=const$ from the contracted Bianchi identity for $R_{ab}$. As the equations are exactly the same as in the mentioned paper, the explicit computations are not repeated here.
Finally, taking the $ij$-component of eq.\eqref{einst1} we get
\footnote{See eqs.(25) and (29) in \cite{Kunduri:2008rs}. Note however that we use slightly different coordinates, which means that their derivatives with respect to $\rho$ are replaced by $x$-derivatives in our case using $\partial_{\rho}=
\sqrt{\det f}\partial_{x}$.}

\ben
R_{ij}=-\frac{
1}{2}[(\det f)(f_{ij})''+(\det f)'(f_{ij})'-(\det f)(f_{ik})'f^{kl}(f_{lj})']=\frac{1}{2}k_{i}k_{j}e^{2\lambda}-\frac{1
}{2}(\det f)(f_{ij})'\lambda'+\Lambda f_{ij}
\een
where $(\ldots)'$ denotes the derivative with respect to $x$.
Raising one index by contraction with $f^{ij}$, multiplying with $e^{-\lambda}$ and rearranging some terms, this equation can be seen to simplify to


\ben
\Big[(\det f)f^{jk}(f_{ik})' e^{-\lambda}\Big]'=-
k_{i}k^{j}e^{\lambda}-2
\delta_{i}^{\phantom{i}j}\Lambda e^{-\lambda}\quad ,
\label{contradict}
\een
where $\delta_{i}^{\phantom{i}j}$ is the Kronecker-delta.
From this equation we want to derive a contradiction to our assumption of non-staticity, mainly using the following basic properties of periodic functions\footnote{We are also assuming our functions to be smooth.}:

\begin{enumerate}
\item The product of two periodic functions with period $P$ is again periodic with period $P$. \label{it1}
\item The derivative of a periodic function with respect to the periodic coordinate is again periodic.\label{it2}
\item The $x$-derivative of a periodic function in $x$ cannot have the same sign for all $x$, which implies that it has to vanish for some $x$.\label{it3}
\end{enumerate}
It follows from items \ref{it1} and \ref{it2} that the left hand side of eq.\eqref{contradict} is the derivative of a periodic function in $x$.
 Therefore, by item \ref{it3} the right hand side has to be a periodic function of alternating sign, and thus has to be vanishing for some $x$. Let us have a look at the off-diagonal entries, i.e. $i\neq j$. Vanishing of the right hand side then implies $k^{j}=0$ or $k_{i}(x)=f_{ij}(x)k^{j}=0$ for some $x$. The latter in turn would yield $\det f=0$ for this value of $x$ and is thus not an admissible solution. It follows that all the constants $k^{i}$ have to vanish, which, however, leads us back to the static case. Hence, no stationary solutions of the type considered exist, and the proof is finished.
\end{prf}
\paragraph{Remark:} The above reasoning actually does not work in $D=4$ dimensions, since in that case there are no off-diagonal entries of $R_{ij}$. However, taking a look at the component $R_{11}$, it is easy to see that the right hand side of eq.\eqref{contradict} is strictly negative for $\Lambda\geq 0$. Thus, the theorem carries over to the 4-dimensional case with non-negative cosmological constant.
\begin{thm}\label{thm2}
There cannot be any smooth, stationary, non-static cohomogeneity-one near horizon geometry with topology $H\cong T^{2}$ satisfying the vacuum Einstein equations with non-negative cosmological constant in $D=4$ spacetime dimensions.
\end{thm}

\section{Conclusions}

We have ruled out the possibility of toroidal stationary, non-static near horizon geometries subject to the vacuum Einstein equations with
 cosmological constant,
 mainly using the fact that the orbit space $H/U(1)^{(D-3)}$ now is a circle instead of an interval. This gives rise to a periodic coordinate $x$ and, as we have shown, no periodic smooth solutions in $x$ obeying our assumptions exist. Our theorem closes a gap in the existing literature, proving e.g. that the classification in \cite{Hollands2009} includes all possible near horizon geometries of the type considered, instead of just the \emph{physically relevant} ones.
 
Possible extensions of our results could include matter fields, like e.q. Einstein-Maxwell theory or supergravity models. It might also be of interest to remove the restriction on the sign of $\Lambda$ in our theorem \ref{thm2}.
\paragraph{Acknowledgements:} I would like to thank S. Hollands for bringing this subject to my attention, for helpful discussions and for proofreading this paper. I would also like to thank the Erwin Schr\"odinger Institute Vienna for its hospitality and financial support during my stays in March and May.

\bibliographystyle{utphys}
\bibliography{bib}

\end{document}